\documentclass[onecolumn,aps,showpacs,floatfix,prc]{revtex4-2}
\usepackage[dvips]{epsfig}
\usepackage{amsmath,bm}
\usepackage{xcolor}[dvips]

\begin{document}

\title{Big Bang: a theory or fact}

\author{D. N. Basu }

\affiliation{Variable Energy Cyclotron Centre, 1/AF Bidhan Nagar, Kolkata 700 064, India }

\email[E-mail: ]{dnb@vecc.gov.in}

\date{\today }

\begin{abstract}

    The discovery and confirmation that some nuclides were formed soon after the Big Bang is one of the strongest arguments in favour of the Hot Big Bang theory. The process of combining protons and neutrons in a hot, expanding universe is known as Big Bang nucleosynthesis (or, occasionally, primordial nucleosynthesis). The only experiment that is currently constructed to be concurrently sensitive to all four known fundamental forces - gravitational, electromagnetic, strong and weak forces - is big bang nucleosynthesis, which offers our earliest test of cosmology. Combined, our theoretical comprehension of Big Bang nucleosynthesis and the measurement of primordial abundances constitute one of the most robust foundations for the conventional cosmological model. This deliberation provides modern calculations of Big Bang nucleosynthesis, help readers gain an intuitive knowledge of the process and give an overview of the most recent state-of-the-art measurements. Our trust in the current basic picture of cosmology is reinforced by the overall amazing agreement between Big Bang nucleosynthesis and many cosmological probes. 

\end{abstract}
   
\maketitle

\noindent
\section{Introduction}
\label{section1}

    The Big Bang model is a theory that states that the universe started out as a singular (or nearly singular) state and then had an expansionary phase that continues to this day. In fact, the Big Bang is the name given to the first event in the universe's history. It is an extrapolation of the universe's current expansion done backward. For this reason, the Big Bang theory does not explain the origin and very beginning of the Universe, despite what is often believed.

    Shortly after the big bang, primordial nucleosynthesis occurred \cite{Ho64}, and the universe swiftly evolved, permitting only the synthesis of the lightest nuclides, such D, $^{3,4}$He and $^{6,7}$Li. During the big-bang nucleosynthesis (BBN), several unstable or radioactive isotopes, such as tritium or ${^3}$H and $^{7,8}$Be, were also generated in addition to these stable nuclei. One of the stable isotopes was created by the decay or fusion of these unstable isotopes with other nuclei. It lasted only about seventeen minutes (during the three to about twenty-minute period from the beginning of space expansion), and after that the universe's density and temperature dropped below the critical values for nuclear fusion, preventing the formation of elements heavier than beryllium while permitting the existence of unburned light components like deuterium. These abundances of nuclides serve as windows into the conditions that existed in the universe at the very beginning of its evolution. It is thought that mainstream models of cosmology, nuclear physics, and particle physics, which determine values of temperature, nucleon density, expansion rate, neutrino content, neutrino-antineutrino asymmetry, etc., adequately explain the conditions during the BBN. Variations from the BBN constrain nonstandard physics or cosmology that could change the conditions during BBN and test the parameters of these models \cite{Wa67,Wa69,Io09}. Previous research has extensively examined the BBN model's sensitivity to various parameters and physics input \cite{No00,Cy02,Cy04,Se04,Fu09}. 
    
\noindent
\section{The Big Bang Model}
\label{section2}

    Three indicators of the big bang concept are the Hubble expansion of the universe, the Cosmic Microwave Background Radiation (CMBR) and the big bang nucleosynthesis (BBN). There are numerous observational evidences that support these. According to \cite{Ho64}, the BBN predicts the primordial abundances of light elements like D, $^{3,4}$He, and $^{6,7}$Li, whose syntheses occurred seconds after the big bang and only permitted the nucleosynthesis of the lightest nuclei thereafter due to the fast growing cosmos. Some unstable radioactive isotopes, such as ${^3}$H and $^{7,8}$Be, were also synthesized during the BBN in addition to the stable nuclei D, $^{3,4}$He, and $^{6,7}$Li. Eventually, through decay or nuclear fusion, these unstable isotopes changed into stable isotopes. Between three and twenty minutes, from the start of space expansion, BBN persisted. After that, the universe's temperature and density dropped below the critical point for a nuclear fusion reaction to occur, preventing the primordial synthesis of elements heavier than beryllium while also preserving unburned light elements (like deuterium) that did not undergo fusion. 
    
\begin{figure*}[h!]
\vspace{0.0cm}
\eject\centerline{\epsfig{file=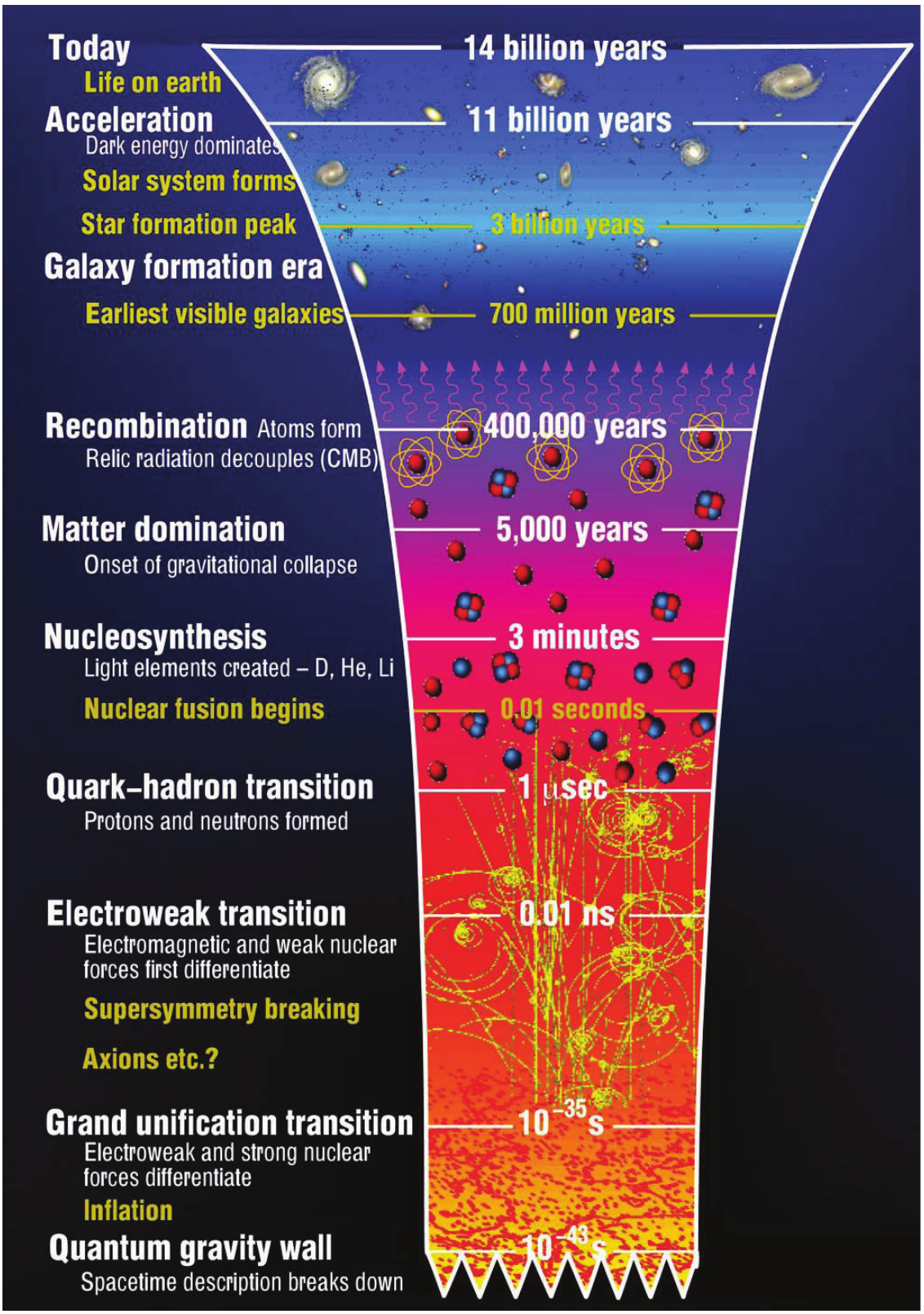,height=20cm,width=14cm}}
\caption{ The history of our Universe. The vertical time axis is not linear in order to show early events on a reasonable scale. (Photo Courtesy: www.ctc.cam.ac.uk)}
\vspace{0.0cm}
\end{figure*}

    The standard BBN theory's predictions are dependent on low energy nuclear cross sections as well as three additional parameters: the universe's ratio of baryons to photons ($\eta=n_B/n_\gamma$), the lifetime of neutrons ($\tau_n$), and the number of flavors of light neutrinos ($N_\nu$) \cite{Co95,Sh95}. Primordial nucleosynthesis is thus a parameter free theory in its typical $N_\nu=3.0$ version, since $\tau_n= 877.75\pm0.28_{stat}+0.22/-0.16_{syst}$ s \cite{Go21} is a precisely measurable quantity and the observations of the CMBR's anisotropies have provided exact information about $\eta=\eta_{10}\times 10^{-10}=6.0914 \pm 0.0438 \times 10^{-10}$ ~\cite{WMAP,WMAP1}. The synthesis of light components in the early cosmos is one of the most important predictions of the mainstream big-bang hypothesis. Big-bang cosmology is based on three main pillars: the BBN, CMBR and the Hubble expansion. Unlike the other two, the BBN looks deeper into the early cosmos and must also address particle and nuclear physics in addition to cosmology. 
    
    The most crucial inputs for modeling the BBN and stellar evolution are the nuclear reaction rates $<\sigma v>$ in reaction network calculations, where $v$ is the relative velocity between the participating nuclides and $\sigma$ is the nuclear-fusion cross section. The cross section $\sigma$ can only be derived by laboratory experiments, some of which are not as well known as expected, even though $v$ is well represented by a Maxwell-Boltzmann velocity distribution for a given temperature $T$ \cite{No00,Cy02,Cy04,Se04}. The measured cross section values are affected by a number of variables, and the different approximations that are employed affect the theoretical estimations of the thermonuclear reaction rates. Any discrepancy \cite{Fo88,An99} in these rates can alter the description of elemental synthesis in the BBN or in stellar evolution, and one must take this into account in the network computations. Based on the BBN approach, the simplified time-line of the Universe's evolution is displayed in Fig.-1. 
    
\subsection{The early Universe — Its time evolution}

    The infinite density and temperature of the universe was the first cosmic singularity. As a matter of fact, we are now able to attempt to characterize the universe in terms of time after the Big Bang, or the so-called Planck time $t_{Pl} \approx 5.39 \times 10^{-44} s$. The temperature and the time evolution of the baryonic density must be known in order to do the nucleosynthesis computation. These can be computed using the universe's expanding rate and thermodynamic considerations. The universe's geometry is described by the Friedmann-Robertson-Walker (FRW) metric, which is provided by  

\begin{equation}
 ds^2 = dt^2 - a^2(t) \Big( \frac{dr^2}{1-k r^2} + r^2(d\theta^2 + sin\theta d\phi^2)  \Big).
\label{seqn1}
\end{equation}
\noindent
It assumes homogeneity and isotropy. The scale factor $a(t)$ characterizes the expansion, and k = 0, $\pm$1 represents the flat, closed, or open universe, accordingly. From Einstein equations one obtains

\begin{equation}
 H^2(t) = \Big(\frac{\dot a}{a}\Big)^2 = \frac{8\pi G (\rho_M + \rho_R)}{3} - \frac{k}{a^2} + \frac{\Lambda}{3}
\label{seqn2}
\end{equation}
\noindent
where $\Lambda$ is the cosmological constant, $\rho_M$ and $\rho_R$ are the matter and radiation densities, respectively, and $H(t)$ is the Hubble parameter. For the density components of the universe, it is convenient to take into consideration the critical density $\rho_C=\frac{3H_0^2}{8\pi G}$ for a flat (Euclidean) space corresponding to $k = 0$ and $\Lambda = 0$ in Eq.(2).

    It is commonly known that in the early phases of expansion, the radiation density $\rho_R \propto a^{-4}$ and the matter density $\rho_M \propto a^{-3}$ (dark and baryonic). When $a$ is roughly $10^{-8}$ times the current value during the BBN era, relativistic particles alone determine $H(t)$; the cosmological constant, curvature terms, and matter density play no part. In this case Eq.(2) takes the form

\begin{equation}
 \Big( \frac{\dot a}{a}\Big)^2 = \frac{8\pi G }{3}a_R \frac{g_*(T)}{2} \times T^4    
\label{seqn3}
\end{equation}
\noindent
where $g_*$ is the effective spin factor, which decreases as and when the temperature drops below a threshold mass for the annihilation of every species of particle with its antiparticle, and the Stefan-Boltzmann formula $a_R T^4$ for the radiation energy density is used. Photons, neutrino/anti-neutrino and e$^+$/e$^-$ contribute to $g_*(T)$ before they annihilate, as only electrons and positrons annihilate during BBN. Because it happens after they disconnect, the energy released is shared by those particles that were in equilibrium with baryons and photons but not with neutrinos. The neutrino temperature, photon/ion temperature, and baryonic density can be obtained as functions of time by solving Eq.(3) \cite{Wa67,Wa69} numerically and applying the constraint that the entropy densities of neutrinos and photons+electrons stay separately constant during the adiabatic expansion \cite{Ka92,We08}. These are the crucial inputs, together with the thermonuclear reaction rates, needed for BBN computations. The baryonic density at this time has no effect on the universe's rate of expansion. On the other hand, a higher baryonic density affects nucleosynthesis by causing a greater number of nuclear reactions per unit time.

    Primordial nucleosynthesis, or BBN, deals with nuclear physics, particle physics, and cosmology and dates back to the beginning of the universe. Even if there are alternative cosmological models that can account for the Hubble expansion, the evidence from CMBR and BBN measurements point to an early cosmos that was extraordinarily hot and dense. The standard scenario that forms the basis of BBN theory is the FRW cosmological model. This approximation is validated by the fact that the standard BBN theory works and that the solution to Einstein's equations results in a homogeneous and isotropic cosmos, implying uniformity of the CMBR temperature across the sky, which is $T= 2.7277 \pm 0.002$ K. The big-bang expansion rate, H, and the thermal characteristics of the particles in existence at that time are related by the Friedmann equation. In nuclear processes, the weak interaction converts neutrons into protons and vice versa, while the strong interaction creates complex nuclei. During expansion, these processes' rates are involved. The duration of the universe's expansion determines the light element syntheses that occurred in its early stages. 

\subsection{Cosmological inflation} 

    The theory that most closely matches data at the moment is that the universe is rapidly expanding exponentially throughout the early stages of its evolution. The entire mechanism is assumed to be driven by a scalar field (inflaton) with self-interaction potential according to the standard approach. The so-called slow-roll approximation, which makes the assumption that the inflaton's potential energy is significantly greater than its kinetic energy, is the most widely used method for scalar field inflationary models. The observational evidence (such as those associated with CMBR) matches extremely well to the typical inflationary model with a single scalar field. It makes a fairly accurate prediction about how the universe's large-scale structure came to be. These factors have led to the cosmological inflationary model becoming somewhat of a paradigm in physical cosmology today. The expansion of the cosmos persisted after the inflationary era, albeit more slowly. Dark energy caused the universe to slow down, but it started to accelerate again after the universe was more than 7.7 billion years old (5.4 billion years ago).
 
\subsection{Baryogenesis and Nucleosynthesis}

    As matter and antimatter should have been formed in equal numbers, the overall baryon number should ordinarily be zero as it is assumed that the particles we see were created using the same physics we measure today. But matter outweighs antimatter in the cosmos. The physical process that is assumed to have occurred in the early universe to produce baryonic asymmetry - that is, the imbalance of matter (baryons) and antimatter (antibaryons) in the observed world - is known as baryogenesis in physical cosmology. The two primary ideas are the GUT baryogenesis, which would occur at or soon after the grand unification epoch, and the electroweak baryogenesis (Standard Model), which would occur during the electroweak phase transition. Mechanisms of this kind are described using statistical physics and quantum field theory. Primordial nucleosynthesis, which is when atomic nuclei started to form, comes after baryogenesis. 

    The reactions which took place during BBN can be organized into two categories, {\it viz} the nuclear reactions which convert neutrons to protons and {\it vice versa}: n $\leftrightarrow$ p $+$ e$^-+\bar\nu_e$; n $+$ e$^+ \leftrightarrow$ p $+~\bar\nu_e$ and p $+$ e$^- \leftrightarrow$ n $+~\nu_e$ and the rest of the other reactions. The first categorization is in terms of the mean lifetime of neutrons while the second categorization is on the basis of cross section measurements of different nuclear reactions. 
    
    The deuterium formation starts with the p $+$ n $\leftrightarrow$ D $+~\gamma$  process. This is an exothermic reaction releasing 2.2246 MeV of energy. The photons are 10$^9$ times more numerous than protons, therefore photo-destruction rate is more than the production rate of deuterons. Hence the deuterium formation reaction can proceed only when the temperature drops to about 0.3 MeV by the expanding universe. The reactions that make ${^4}$He nuclei: D$~+ n \rightarrow {^3}$H$~+ \gamma$, ${^3}$H$~+~$p $\rightarrow {^4}$He $+$ $\gamma$, D $+$ p $\rightarrow {^3}$He $+~ \gamma$, ${^3}$He $+$ n $\rightarrow {^4}$He $+~\gamma$  follow once the deuteron formation starts. Apart from the ${^4}$He (normal helium), ${^3}$He (light helium) is also formed along with the ${^3}$H. The ${^4}$He nucleus has a binding energy of 28.3 MeV and is more bound than the deuterons. Moreover, the temperature of the plasma by this time has dropped down to 0.1 MeV already, forces these reactions (being photo-reactions) to proceed one way only. The reactions: D $+$ D $\rightarrow {^3}$He $+$ n, D $+$ D $\rightarrow {^3}$H + p, ${^3}$He $+$ D $\rightarrow {^4}$He $+$ p, ${^3}$H $+$ D $\rightarrow {^4}$He $+$ n also produce ${^3}$He and ${^4}$He. These four reactions are not associated with the relatively slow process of emission of photons and hence usually proceed faster. The reaction between deuterons and other charged particles eventually stops due to electrostatic repulsion when the temperature falls too low to overcome it. When these reactions stops, the ratio of deuteron to proton is very small as it varies as -1.6 power of the total density of neutrons and protons. Most of the neutrons of the universe at that epoch end up as ${^4}$He nuclei. At the time of deuteron formation, neutron:proton ratio is about 1:7 and 25$\%$ of the mass ends up in the synthesis of the helium. After about 100 seconds of the big-bang the deuterium concentration peaks which subsequently ends up in helium nuclei. Thereafter, a very small amount of helium nuclei can fuse to form heavier nuclei producing a small BBN ${^7}$Li abundance. As half-life of ${^3}$H which decays into ${^3}$He is twelve years, so no primordial ${^3}$H can survive till now. Similarly, half-life of ${^7}$Be which decays into ${^7}$Li is fifty three days, so no primordial ${^7}$Be also can exist today.  The uncertainties involved in the reaction rates of ${^3}$He$+{^4}$He $\rightarrow {^7}$Be$+\gamma$, ${^3}$H$+{^4}$He $\rightarrow {^7}$Li$+\gamma$ and $p+{^7}$Li $\rightarrow {^4}$He$+{^4}$He reactions could result in a projected yield of ${^7}$Li obtained using BBN code \cite{Ka92} that could be uncertain by an amount as high as nearly 50$\%$. In Fig.-2, a typical Big Bang Nucleosynthesis nuclear reaction network has been shown. The bold arrows indicate which twelve reactions are most significant.
    
    Instead of cross sections $\sigma$, the thermonuclear reaction rates are used as the inputs to BBN calculations. These rates are calculated by folding Maxwell-Boltzmann energy distribution with energy dependent nuclear reaction cross sections. Thus the Maxwellian-averaged reaction rate per interacting particle pair $<\sigma v>$ at a temperature $T$, can be described by the equation \cite{Bo08,Ad11} given below:

\begin{equation}
 <\sigma v> = \Big[\frac{8}{\pi\mu (k_B T)^3 } \Big]^{1/2} \int \sigma(E) E \exp(-E/k_B T) dE,
\label{seqn4}
\end{equation}
\noindent
where the Boltzmann constant, the energy in the center-of-mass frame, the reduced mass of the reactants and the relative velocity are represented by the variables $k_B$, $E$, $\mu$ and $v$, respectively. The radius of the nucleus is too tiny compared to the classical turning point at energies significantly below the Coulomb barrier. With $e$ representing the elementary charge and $Z_1$ and $Z_2$ representing the atomic numbers of the interacting nuclei, $\eta = \frac{Z_1Z_2e^2}{\hbar v}$ is the Sommerfeld parameter in this case, the quantity $\exp(-2\pi\eta)$ roughly approximates the tunneling probability through the barrier. The charge-induced cross section is therefore factorized as

\begin{equation}
 \sigma(E) = \frac{S(E)\exp(-2\pi\eta)}{E}.
\label{seqn5}
\end{equation}
\noindent
Insofar as the narrow resonances are eliminated, $S(E)$, the astrophysical $S$-factor, is a smooth function of energy, making it easier to extrapolate the empirically measured cross sections down to the energies of the astrophysical realm. While the cross sections for neutron-induced reactions are given at low energies by $\sigma(E)=\frac{R(E)}{v}$ \cite{Bl55}, for the narrow resonance case, a Breit-Wigner expression generally approximates the resonant cross section. Here, $R(E)$ is a function that varies slowly with energy \cite{Mu10} and is comparable to the $S$-factor. The best test of the Standard Model for the first few minutes following the Big Bang is the combined examination of several measures of primordial abundance, intricate computations and laboratory measured nuclear reaction rates.

\subsection{Cooling of the universe}

    The universe's density and temperature are both still declining. Phase transitions that break symmetry are probably going to happen, which will separate fundamental interactions from a single unified interaction at first. Annihilation results in the remaining matter particles only because of the asymmetry between matter and antimatter. The CMBR is a remnant of the electron and proton recombination era that is released about 0.38 million years after the Big Bang when the plasma cooled to 3000 Kelvin. The photons then dispersed out into space and had a free course. The CMBR appears to be a crucial piece of evidence supporting the Big Bang theory of the universe's creation. 

\subsection{Structure formation in universe}

    Structure formation in physical cosmology refers to the emergence of galaxies, galaxy clusters and bigger structures that originate from minute fluctuations in mass density brought about by matter-creating processes. Areas of uniformly distributed matter that are a little denser will gravitationally draw in surrounding matter, becoming denser still. Stars, galaxies, gas clouds, and other celestial formations that are visible to us today are produced. The process of re-ionization, which happened between 150 million and 1 billion years after the Big Bang, is significant at this phase. The observed large-scale distribution of galaxies, clusters, and voids can be accurately predicted by the current Lambda-CDM model; however, there are numerous complications at the individual galaxy scale because of highly nonlinear processes involving gas heating and cooling, star formation, feedback, and baryonic physics. One of the main focuses of current cosmology research is comprehending the processes of galaxy formation, using both massive computer models and observations like the Hubble Ultra-Deep Field. 

\clearpage
{\bf The BBN Network Diagram}

\begin{figure*}[h!]
\vspace{0.0cm}
\eject\centerline{\epsfig{file=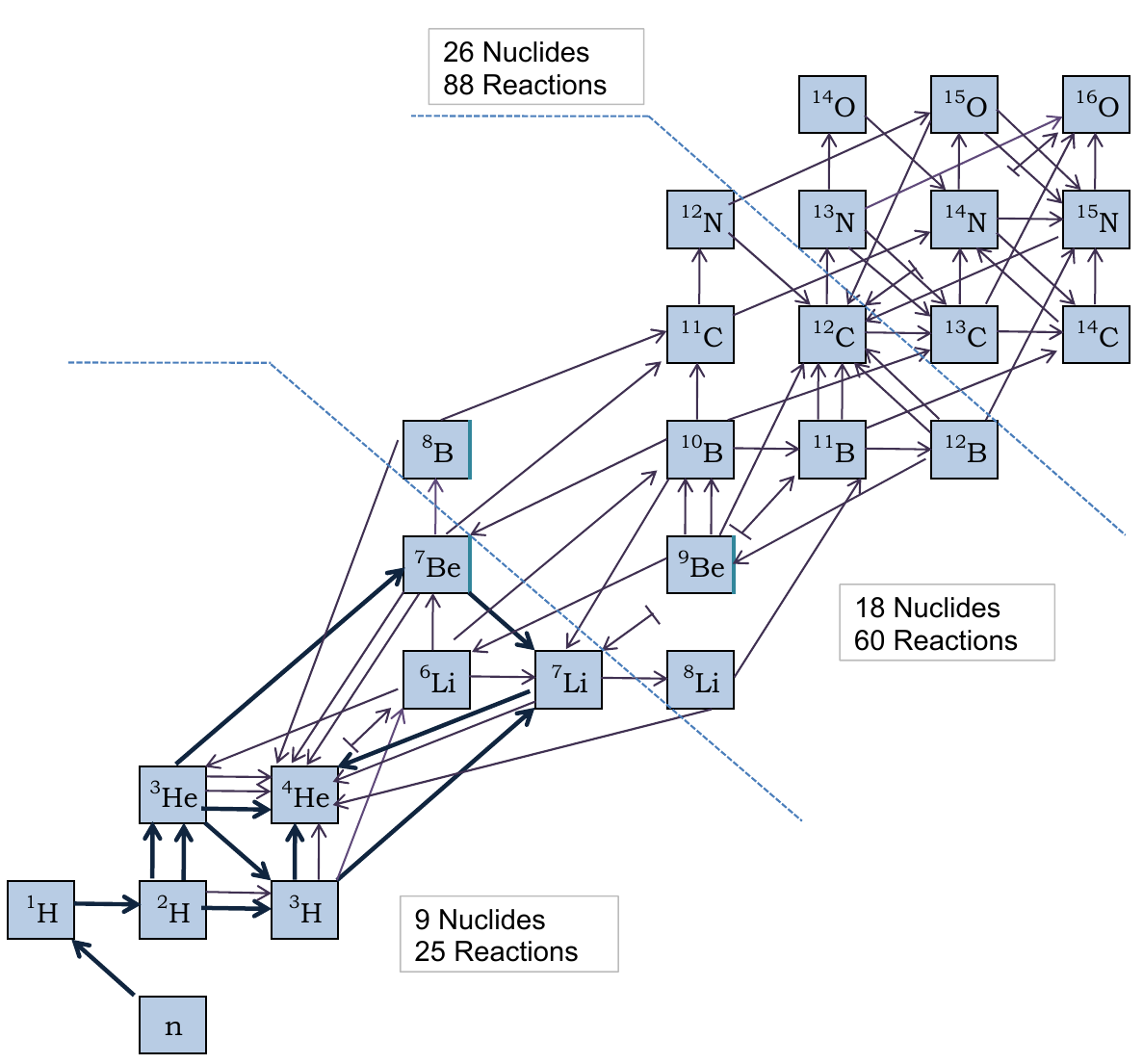,height=14cm,width=14cm}}
\caption{Outline of nuclear reaction network for Big Bang Nucleosynthesis. The twelve most important nuclear reactions which affect the predictions of abundances of light elements have bold arrows.}
\vspace{0.0cm}
\end{figure*}

\subsection{Accelerating expansion}

    The period of accelerating expansion is the period when dark energy predominated and is characterized by negative pressure. With an estimated age of 13.79 billion years, the Universe has been expanding at an accelerated rate for approximately 5 billion years, assuming that the dark energy is a cosmological constant. Nevertheless, the approach presents certain limits in terms of its applicability and associated issues.
    
\noindent
\section{Status of primordial abundances}
\label{section3}

    After BBN, stars also produce ${^4}$He. Its primordial abundance is deduced from observations in compact blue galaxies' ionized hydrogen regions. According to a hierarchical structure creation paradigm, galaxies are believed to be created through the accumulation of these more primordial dwarf galaxies. The ${^4}$He abundance derived from observations is extrapolated to zero, and then atomic physics corrections are applied to account for stellar generation. Its mass fraction was estimated by Aver et al. to be $0.2449 \pm 0.0040$ \cite{Aver15}; however, in 2021, with less uncertainty than the first estimate, it was updated to $0.2453 \pm 0.0034$ \cite{Aver15}.  

    Deuterium can be removed throughout star evolution after BBN. By looking at a few cosmic clouds in the line of sight of distant quasars at high redshift, one can determine the primordial amount of deuterium. A D/H relative abundance of $(2.53 \pm 0.04) \times 10^{-5}$ was obtained from a recent reanalysis of earlier data and additional observations by Cooke et al. \cite{Cooke14}. This estimate was revised to $(2.527 \pm 0.030) \times 10^{-5}$ \cite{Cooke14} in the year 2018 with lower uncertainty than earlier estimate.

    As ${^3}$He is both generated and destroyed in stars, unlike ${^4}$He, it is challenging to estimate how its abundance will vary with time. Due to the difficulties in helium research and the low ${^3}$He/${^4}$He ratio, the relative abundance of ${^3}$He is estimated to be $(<1.1 \pm 0.2) \times 10^{-5}$ \cite{Ba02} based only on the observations in our Galaxy.
    
    The BBN continued three to twenty minutes after space expansion began. Because of this, the universe's density and temperature dropped below that required for nuclear fusion, which allowed unburned light elements like deuterium to exist but hindered the production of things heavier than beryllium. The majority of heavy element nucleosynthesis takes place in big stars. As galaxies evolve, these stars explode as supernovae, releasing elements rich in heavy metals into the interstellar medium. Consequently, with time, stars' abundances of heavier elements increase. Therefore, the measured abundance of metals (elements heavier than helium) can be used to infer the age of the star. As a result, the older stars have less metallicity. Very low metallicity objects were observed in order to derive the primordial abundances. After BBN, $^7$Li can be destroyed (inside the core of stars) as well as formed (by novae, AGB stars and spallation). Very old stars can still be spotted in the halo of our galaxy because stars with masses smaller than the Sun have a lifespan longer than the age of the universe. It was shown that the abundance of lithium is remarkably stable, independent of metallicity, as long as it is less than $\approx$0.1 of solar metallicity. Lithium can be observed at the surfaces of such low metallicity old stars. It was previously believed that the synthesis of $^7$Li in BBN corresponded to this steady plateau in Li abundance. Given the thinness of the plateau, it is possible that surface Li depletion was not very successful, and that the primordial value was instead reflected. Sbordone et al.'s study \cite{Sb10} provides an estimate of $^7$Li/H $=(1.58^{+0.35}_{-0.28}) \times 10^{-10}$.   

\begin{table*}[htbp]
\vspace{0.0cm}
\centering
\caption{\label{tab:table1} Yields at CMB-WMAP baryon-to-photon ratio of Universe ($\eta_{10}=6.0914 \pm 0.0438$ ~\cite{WMAP,WMAP1}) using latest neutron lifetime value $\tau_n=877.75\pm0.28_{stat}+0.22/-0.16_{syst}$ s \cite{Go21}.}
\begin{tabular}{lccl}
\hline
\hline
 
Element & Abundance & Observations & Ref. Year \\ \hline 
${^4}$He&0.2461$\pm$0.0002 &0.2453$\pm$0.0034 &\cite{Aver15}~Aver et al. 2021 \\

D/H~($\times 10^{-5}$)&2.620$\pm$0.031 &2.527$\pm$0.030 &\cite{Cooke14}~Cooke et al. 2018 \\

${^3}$He/H~($\times 10^{-5}$)&1.066$\pm$0.005&$<$1.1$\pm$0.2 &\cite{Ba02}~Bania et al. 2002 \\

${^7}$Li/H~($\times 10^{-10}$)&4.421$\pm$0.067 &1.58$^{+0.35}_{-0.28}$ &\cite{Sb10}~Sbordone et al. 2010 \\

\hline
\hline
\end{tabular} 
\vspace{0.0cm}
\end{table*}
\noindent

    Table-I presents a comparison of results from theoretical computations \cite{Si24} and observations \cite{Aver15,Cooke14,Ba02,Sb10}. The theoretical errors originate from the experimental uncertainties in the values of $\tau_n$ and $\eta$. It may be observed that there is still an overestimation of the $^7$Li abundance, despite the fact that the computed abundances of light elements during primordial nucleosynthesis and those found from observations accord well over a range of nine orders of magnitude. It is worth mentioning that the relative abundance of $^7$Li ($1.1\pm0.1 \times 10^{-10}$), which was originally determined by Hosford et al. \cite{Ho09}, was revised upward by about 44 percent by a subsequent research \cite{Sb10}. The lowest limit of the current theoretical estimate is still higher than the upper limit of the measured value of $^7$Li relative abundance, even though the theoretical and observed values seem to be converging. It is reasonable to opine that there may be room for further improvement considering the fact that the detection of primordial lithium in extremely metal-poor stars within the galactic halo is currently not possible with current observational capabilities.

\noindent
\section{Problems associated with Big Bang}
\label{section4}

\subsection{Initial singularity problem}

    The cosmos expanded from an extremely hot, dense state, as evidenced by the cosmic microwave background, even though there is no direct evidence for a singularity of infinite density. Although an initial singularity in the case of general relativity is unavoidable \cite{Ha70}, it is not predicted by standard inflation theory. As a result, pre-inflationary times are still inaccessible given our current understanding of the cosmos. Some variants of the Big Bang hypothesis suggest that the original singularity existed prior to the Big Bang. The instant that comes right after the first singularity is a portion of the Planck epoch, which is the oldest time in the Universe's history. Because the laws of physics as they exist today break down at the earliest times in the Universe's history, we are unable to determine if the Universe started with an initial singularity. This issue might become clearer once the theory of quantum gravity is fully developed.

\subsection{Asymmetry between matter and antimatter}
    
    Several theoretical explanations are put forward to explain baryonic asymmetry, including the identification of favorable conditions for the formation of normal matter (as opposed to antimatter) and symmetry breaking. A tiny fraction of a second after the Big Bang, of the order of 1 in every 1630000000 ($\approx$2$\times$10$^9$) particles, is required for this imbalance to exist. All of the baryonic matter and a significantly higher number of bosons remained in the current universe after the majority of matter and antimatter were destroyed by annihilation. However, it appears from Fermilab experiments \cite{Fermilab2010} that this imbalance is far more than estimated. In these particle-collision studies, the amount of matter produced was determined to be around 1$\%$ greater than the amount of antimatter produced. This disparity's cause is currently unknown.

\subsection{Horizon problem}

    At cosmological scales, the homogeneity and isotropy of the Universe defy the mainstream physical cosmology. The homogeneity level of $10^{-5}$ makes the CMBR map impossible without an additional physical mechanism to be so homogeneous. It is impossible for photons at the two farthest `edges' of the universe to be causally related.
    
    Solution to the Horizon problem: The observable universe grew out of a small, originally homogeneous zone that rapidly expanded during the inflationary era, making it homogeneous and isotropic on vast scales. Particles could interact with one another prior to cosmic inflation because the homogeneity scale was always greater than the causality scale.

\subsection{Flatness problem}

    The Universe is spatially flat (space is characterized by Euclidean geometry) based on existing observational evidence, but this is only true if the density of the Universe is nearly exactly equal to the so-called critical density (with a precision level of at least $10^{-56}$ ). The spatial geometry (and consequently the evolution of the Universe) would be altered by even a little variation in the value of this parameter.
    
    Solution to the Flatness problem: Unlike matter and radiation, the energy of the scalar field (inflaton) remains relatively constant during the cosmic inflation. If the expansion of the Universe is allowed to continue at an exponential rate, the near-perfect match of the Universe's density to the critical density may be reduced to the current accuracy of 1$\%$.

\subsection{Magnetic monopoles}

    Topological defects are predicted to exist in the spatial structure of the universe by the grand unification theories (GUTs), which aim to unify all fundamental interactions; however, no evidence of such items has been discovered to date.
    
   Solution to the Magnetic monopoles problem: All phase transition remains (topological defects, magnetic monopoles, etc.) are extremely far apart as a result of cosmological inflation that takes place at a temperature (energy) lower than that anticipated by GUTs. They are currently invisible because of their extremely low density.

\subsection{Dark Energy and Dark matter}

    Dark energy seems to be in charge of the universe now. Dark energy makes up around 3/4 of the universe's total mass-energy, while matter, both dark and regular, makes up the remaining 1/4. The radiation from stars and galaxies is far weaker than the radiation from the universe, which makes up only a tiny portion and is largely caused by CMBR. Dark energy is a term used to describe an unidentified type of energy that has the biggest effects on the universe in physical cosmology and astronomy. Its main impact is to propel the universe's accelerated expansion. Supernova measurements provided the first observational proof of dark energy's existence. Because Type 1A supernovae have a consistent luminosity, they are useful for measuring distances precisely. When this distance is compared to the redshift, it becomes evident that the universe is expanding faster. A theorized type of stuff that does not seem to interact with light or the electromagnetic field is known as `dark matter' in astronomy. Gravitational effects suggest the existence of dark matter, which general relativity cannot account for unless there is more matter than what is visible. These effects arise in relation to galaxy formation and evolution, gravitational lensing, the current structure of the observable universe, mass location in galactic collisions, galaxy mobility within galaxy clusters, and cosmic microwave background anisotropies.

   The current mass/energy balance of the universe indicates that dark matter makes up more than 24$\%$ of its total mass whereas and normal matter constitute less than 5$\%$. Current models attribute the gravitational stability of galaxies to this material forming a so-called halo around them. It seems to just be in contact with gravity. The Standard Model does not apply if this is a novel kind of particle. The universe is currently expanding faster than ever thanks to a component that makes up about 71.4$\%$ of its present content and is characterized by negative pressure. Though nature of dark energy is even more enigmatic than that of dark matter, there are various theories regarding it, including the possibility that it has a constant value or changes over time.
   
\noindent
\section{Anticipations regarding the immediate future, an outlook}
\label{section5}
 
    At this particular point in the history of human cognition, the picture of the elements' origins is starting to become clearer. There is  unanimity that in the nucleosynthesis of the elements during the universe's evolution there have been three stages. The genesis of the universe is connected to the first phase, referred to as the standard Big Bang nucleosynthesis (SBBN). The recombination period marks the start of the second stage, which extends until the stelliferous era, which coincides with the end of the dark ages and marks the appearance of light sources with the arrival of the first generation stars.
    
    The first generation stars created a great deal of new elements, all the way up to the iron peak, in addition to visible light. There are still many unanswered questions regarding the typical mass scale of a first generation star, despite the general consensus that the thermonuclear processes inside stars have produced all elements between C and the Fe peak, either during the stars' quiet evolutionary stages or during the catastrophic explosions (supernovae) that signal the end of lives of some stars.
    
    With reference to the normal mass scale of the first generation stars, there has been a gradual but noticeable paradigm shift. The belief in the historical existence of stars with masses exceeding 100 solar masses gave way to computer simulations that produced stars with masses significantly smaller than that of a few tens of solar masses. With the launch of the James Webb Space Telescope (JWST) in 2021, this mystery may soon be resolved. In the not too distant future, these first generation stars would very possibly be detected by the JWST if they were in fact such incredibly massive objects. Furthermore, if it turns out that first generation stars were big, this will identify them as the progenitors of the supermassive central black holes that exist in galaxies today.
    
    Future forecasts in nuclear physics must be in line with limitations placed on the abundances of heavy elements, which are determined from astronomical measurements. The universe itself, the ultimate laboratory, will be used to test these predictions. In this sense, nature has been generous to scientists by enveloping the Milky Way in a halo of small ultra-faint dwarf satellite galaxies that contain chemically primitive stars, particularly old and low metallicity stars that show large quantities of r-process elements. Though research on this class of stars is still in its early stages, it should yield some illuminating answers on the sources of its r-process elemental abundances. However, the light and delicate isotopes of Li, Be, and B are created by spallation reactions, in which high-energy cosmic ray particles eliminate nucleons from the interstellar medium's plentiful C, N and O nuclei, rather than by star interior production (where their elimination is caused by elevated temperatures).

    The LIGO/Virgo collaboration discovered the second event in 2019 on merging neutron stars of binaries. The second of many multi-messenger observations to come, hopefully, is this finding. These gravitational wave discoveries, along with upcoming extremely precise electromagnetic observations made with both space-based and ground-based detectors, may provide answers to long-standing queries regarding the equation of state for supranuclear matter and the interior structure of neutron stars. In this regard, the International Space Station now has a payload dedicated to the study of neutron stars using soft X-ray timing. June 3, 2017, was the most recent launch date for this mission. It is anticipated that this mission, known as the Neutron star Interior Composition Explorer, would provide some insight into the fundamental characteristics of neutron stars.
    
    Finally, the complete history of the cosmos depends heavily on the accurate determination of the elements in the periodic table. The abundances of the initial elements and their isotopes (H, He, Li, and D) define the SBBN scenario's early stages, including the number of relativistic species existent during that time, the rate of universe expansion, and the nuclei's reaction rates. Measurements of the acoustic peaks of CMBR at last scattering surface, which occurred just 0.38 million years after the Big Bang, can be used to finely restrict details from SBBN. These days, anisotropies of the CMBR at various sky angles are used to measure these acoustic peaks. These are highly dependent on the quantity of baryons (protons and neutrons) in the universe, and to a lesser extent, on the amounts of lithium and helium. Fields et al. \cite{Fields20} reported that the amount of baryonic matter was determined with an accuracy of 0.9$\%$ using data from the Planck spacecraft. This is also related to the number of lepton families (relativistic neutrinos) that the CMBR data allows, which is reasonable, given the abundances of lithium and helium.
    
    In around ten years from now, more refined CMBR probes, referred to as CMB-Stage 4, will provide the information needed to validate or refute existing scientific theories. The lithium problem, which is still present and presents a significant obstacle to theoretical nuclear reaction rates and data consistency, is one of the fundamental problems. Additionally, the number of relativistic degrees of freedom (neutrino families) and the total mass of neutrinos will be able to be determined in the near future through data analysis of CMBR probes with those of some new galaxy surveys, such as the Dark Energy Spectroscopic Instrument jointly. This will help to constrain the helium abundance in the universe and the number of neutrino families. 
    
\section{ Summary and conclusion }
\label{section6}

    Predicting the primordial elemental abundances in the BBN is one of the three arguments in favour of the big bang hypothesis. Because of the precise understanding of the universe's baryon-to-photon ratio gleaned from CMBR anisotropy investigations, the SBBN is a parameter-free theory. Primordial abundances are determined by the number of light neutrino flavors, the neutron lifetime, the baryon-to-photon ratio and the rates of astrophysical nuclear reactions. The theoretically calculated $^7$Li abundance is nevertheless exaggerated, even if the computed abundances of light elements during primordial nucleosynthesis and those found from observations accord well over a range of nine orders of magnitude.
    
     According to current thinking, the `lithium problem' mismatch could result from systematic mistakes in the known abundances, uncertainties in stellar astrophysics, such as nuclear inputs or stellar depletion, or even novel physics that goes beyond the Standard Model \cite{Fields11}. A detailed account of physics up to the Fermi scale can be found in the Standard Model, but cosmology cannot be tracked down before the Big Bang.    
    
    Big Bang Nucleosynthesis now shows where the line is drawn between accepted and theoretical cosmologies. Since no remnants of these eras have been found to far, it is currently unknown how to push this frontier back to the quark-hadron transition or electroweak symmetry breaking.
    
    We have made significant progress in our knowledge of the cosmic origin of the elements in the second decade of the twenty-first century. The coming years are predicted to be a time of collaborative probe analysis, combining data and improving theoretical parameter constraining by measuring light from the CMBR, from galaxy clustering, from individual astronomical objects, and, last but not least, from gravity waves originating from the collapse of neutron stars. New ground and space-based observatories as well as the next generation of nuclear accelerators are on the horizon. Perhaps the current queries will have their answers, but new ones will undoubtedly surface.
    
\begin{acknowledgments}

    The author acknowledges support from Anusandhan National Research Foundation (erstwhile Science and Engineering Research Board), Department of Science and Technology, Government of India, through Grant No. CRG/2021/007333.

\end{acknowledgments}

\end{document}